\documentclass[aps,prd,reprint,showkeys,showpacs,nofootinbib]{revtex4-1}

\usepackage[dvipdfmx]{graphicx} %
\usepackage{amsmath} %
\usepackage{amssymb} %
\usepackage{float} %
\usepackage{color} %

\begin{document}

\title{Generalized framework for testing gravity with gravitational-wave propagation. II. Constraints on Horndeski theory}

\author {Shun Arai${}^1$}
\email{arai.shun@a.mbox.nagoya-u.ac.jp}

\author {Atsushi Nishizawa${}^{2}$}

\affiliation {${}^1$Department of Physics and Astrophysics, Nagoya University, Nagoya 464-8602, Japan}
\affiliation {${}^{2}$Kobayashi-Maskawa Institute for the Origin of Particles and the Universe, Nagoya University, Nagoya 464-8602, Japan\\}

\date{\today}

\begin{abstract}
 Gravitational waves (GW) are generally affected by modification of a gravity theory during propagation in cosmological distance. We numerically perform a quantitative analysis on Horndeski theory at cosmological scale to constrain the Horndeski theory by GW observations in model-independent way.
 We formulate a parameterization for a numerical simulation based on the Monte Carlo method and obtain the classification of the models that agrees with cosmic accelerating expansion within observational errors of the Hubble parameter. As a result, we find that a large group of the models in the Horndeski theory that mimic cosmic expansion of $\Lambda$CDM model can be excluded from the simultaneous detection of a GW and its electromagnetic transient counterpart.  Based on our result and the latest detection of GW170817 and GRB170817A, we conclude that the subclass of Horndeski theory including arbitrary functions $G_4$ and $G_5$ can hardly explain cosmic accelerating expansion without fine-tuning.

 
\pacs{xxxxx}

\end{abstract}

\maketitle

\section{\label{sec:introduction}Introduction}
As is already known, the universe expands acceleratingly. From the first direct measurement of the cosmic accelerating expansion with type-Ia supernovae \cite{Riess,Perl}, the following observations such as cosmic microwave background (CMB) and surveys of large scale structure (LSS) have strongly suggested that Lambda-Cold-Dark-Matter ($\Lambda$CDM) model is the best explanation to describe the dynamical evolution of the late-time universe \cite{WMAP7yrs, Planck2015}. However, the $\Lambda$CDM model is less supported from the theoretical point of view, because it has to assume the existence of unknown components over 95 $\%$ of the total in the universe. Therefore, the cosmic accelerating expansion remains one of the largest riddle of modern cosmology. 

To solve this problem, there are mainly two ways proposed; the dark energy and modified gravity. The former one modifies only energy components of the universe, remaining gravity described with General Relativity (GR). For instance, quintessence \cite{Peebles-Ratra2003} and the nonlinear kinetic term of a scalar field \cite{Chiba2000,K-essence} are the models in the former category. The latter one, on the other hand, prescribes the cosmic accelerating expansion with modification of gravity at cosmological scales. 
 Currently, in the case of modification with a scalar field, the universal description to treat dark energy and modified gravity together is called scalar-tensor theory.  In particular, Horndeski theory  \cite{Horn1974, Ginf2011} is the most general form of theories with a space-time curvature and a scalar field whose equation of motions contain up to second order of space-time derivatives. The Horndeski theory includes not only the quintessence and nonlinear kinetic theory, but also many specific theories; $f(R)$ theories \cite{f(R)theories}, covariant galileons \cite{CovGal,Galileon}, and kinetic gravity braiding \cite{Braiding}.  The Horndeski theory can be also extended further to a more general framework in the language of an effective field theory (EFT). EFT for dark energy is formulated by Gubitosi et al. \cite{EFTofDE}, Gleyges et al.\cite{BldgDE,intDE}, and Bellini and Sawicky \cite{Maximal2014}.
 
With the EFT formulation, some alternative theories against the $\Lambda$CDM model are examined by using CMB and LSS observations \cite{HornCMB}. However, one still needs to specify a model  when calculating observables and compares with observational data. This makes the analyses model-dependent and may cause to overlook a sort of theories that is not classified into the specific models. Therefore, it is important to investigate the whole subclasses in the Horndeski theory in a numerical way beyond analytical difficulties. 

In the meantime, gravitational waves (GW) are gathering attention as a new tool to probe the modified gravity. In fact, LIGO has accomplished to capture GW for the first time \cite{Abbott:2016blz}.  For now LIGO has successively confirmed three more detection from the coalescence of binary black holes (BBH) \cite{GW151226,GW170104, GW170814} at cosmological distance.  Moreover, the event rate of detection of GW with the second-generation detector network at design sensitivity is expected to be $100 - 1000 {\rm yr}^{-1}$, which is statistically enough for cosmological applications. As a result, we are able to extract cosmological information from the GW observations. For instance, GW are applicable to measure the Hubble constant \cite{Schutz1986, Nissanke2013,AN2016, GW170817_Hubble}. 

For the application to constrain on modified gravity that explains the cosmic accelerating expansion, GW are a promising tool. In fact, Lombriser and Taylor \cite{Lombriser2016} and Bettoni et al \cite{Dario2016} have reported that GW detection potentially distinguishes the models of the Horndeski theory filed describing the cosmic accelerating expansion. According to this paper, the detection of the phase velocity of GW, $c_T$, can exclude the wide range of the models that realize the accelerating expansion because the deviation of $c_T$ from GR happens at the same time. Indeed, the measurement of $c_T$ can reach down to $|c_T - 1| \lesssim 10^{-15}$ with GW Cherenkov radiation \cite{Cherenkov_cT} or electromagnetic counterparts for GW emission \cite{AN_TN_cT}. With these remarkable features, we will expect GW to be a tool to search for the nature of the cosmic accelerating expansion.

In this paper, we develop the previous result in Lombriser and Taylor \cite{Lombriser2016} to a numerical approach to test all possible models of the Horndeski and analyze the feature imprinted on propagation of GW. This paper is organized as follows. First of all, we review the previous study on modification of GW waveform during propagation in Sec.~\ref{sec:GW}.  Next, in Sec.~{\ref{sec:Horn_rp}} we construct a model independent method of parameterization to distinguish models in the Horndeski theory at cosmological scales. In Sec.~\ref{sec:Horn_MC}, we compare with the analytical result in \cite{Lombriser2016} a more qualitative way with Monte Carlo simulations. Then we found considerable deviation from their result.  Finally, in Sec.~\ref{sec:real_obs} we obtain the distributions of models on the parameter plane, indicating that GW are the most efficient  probe to constrain the models of modified gravity.

Very recently, LIGO and VIRGO have detected a binary neutron star (BNS) merger named GW170817 \cite{GW170817}. This event is special because a few of gamma-ray telescopes simultaneously caught the signal of a short gamma-ray burst, GRB170817A, and it was identified as the electromagnetic transient counterpart of GW170817. By using the difference of the arrival times between GW170817 and GRB170817A, they obtained a stringent constraint on $c_T$ down to $-3.0 \times10^{-15} \lesssim c_T - 1 \lesssim 7.0 \times10^{-16}$ \cite{GW170817_cT}. Combining this result and our report, we conclude that the models in the Horndeski theory including the arbitrary functions $G_4$ and $G_5$ are difficult to account for the cosmic accelerating expansion. We will discuss this in Sec.~\ref{sec:real_obs}.
  
\section{\label{sec:GW}Modification of Gravitational Waves propagation in cosmological scale}
We briefly introduce how GW is deformed  during propagation because of the modification of gravity. This argument has originally proposed in Saltas et al. \cite{Saltas2014} and recently it has been extended by Nishizawa to a general framework to test gravity theories \cite{AN_testGravity_GWpropagation}. In these papers, the propagation equation of GW is generally given by
\begin{align}
h_{ij}'' + (2+\nu){\cal H}h_{ij}' + (c^2_T k^2 + a^2\mu^2) h_{ij} = a^2\Gamma \gamma_{ij}\,,\label{mod_GW}
\end{align}  
where $h_{ij}$ is a tensor perturbation (GW) and $'$ denotes the derivative with respect to conformal time. In the Eq.~(\ref{mod_GW}) there are four time dependent parameters $\nu$, $c_T$, $\mu$, and $\Gamma$.
$\nu$ is the Planck mass run rate, $c_T$ is the phase velocity of a GW and $\mu$ is graviton mass. $\Gamma$ denotes extra sources generating GW. In a case that $\nu$, $c_T$, $\mu$ are slowly-varying functions with cosmological time scale and there is no source i.e. $\Gamma = 0$, the solution of Eq.~(\ref{mod_GW}) is given in \cite{AN_testGravity_GWpropagation} as
\begin{align}
h = {\cal C}_{\rm MG}h_{\rm GR}\,, \label{h}
\end{align}
where
\begin{align}
&{\cal C}_{\rm MG} \equiv e^{-\cal D}e^{- ik\Delta T}\,,\label{C_MG}\\
&{\cal D} \equiv \frac{1}{2}\int^\tau{d\tau'\nu{\cal H}}\,,\label{calD}\\
&\Delta T \equiv \int^\tau{d\tau'\left \{ (1-c_T) - \frac{a^2\mu^2}{2k^2} \right \}}\,.\label{Delta_T}
\end{align} 
${\cal D}$ and $\Delta T$ correspond to amplitude damping index and additional time delay of GW respectively. We see the damping parameter $\nu$ only appears in GW amplitude, while $c_T$ and $\mu$ are both involved in GW phase. In the Horndeski theory expressed in  $\alpha$ parametrization \cite{Maximal2014}, we translate $\nu$ and $c_T$ into $\alpha$ functions as
\begin{align}
&\nu = a_M\,,\label{nu}\\
&c^2_T = 1+\alpha_T\,. \label{cT}
\end{align} 
and the graviton mass is massless, $\mu = 0$.



\section{\label{sec:Horn_rp}Numerical formulation of Horndeski theory at cosmological scale}
In this section we provide a numerical formulation of Horndeski theory independent of specific models.
As we mentioned in Sec.~{\ref{sec:introduction}}, the current observational analyses for the Horndeski theory are only limited to specific models. This is because the Horndeski theory is too general to investigate. Consequently, one can hardly extract information for which models are relatively favored from others under observational constraints. To tackle this difficulty, we perform a Monte Carlo simulation with a suitable parameterization independent of specific models. Although the number of parameters would become large, numerically this is not a problem. Now we construct a parameterization in the Horndeski theory in the following way. First of all, we give a general Lagrangian density\footnote{$G_2 (\phi,X)$ is often written $K(\phi, X)$ in literature.} of the Horndeski theory as
\begin{align}
{\cal L} = \sum^5_{i=2} {\cal L}_i\,.\label{fullLag}
\end{align}
where
\begin{align}
&{\cal L}_2 = G_2(\phi, X)\,,\label{L2}\\
&{\cal L}_3 = -G_3(\phi,X)\Box \phi\,, \label{L3}\\
&{\cal L}_4 = G_4(\phi, X)R + G_{4X}(\phi, X)\left[(\Box \phi)^2- \phi_{;\mu\nu}\phi^{;\mu\nu} \right]\,,\label{L4}\\
&{\cal L}_5 = G_5(\phi, X)G_{\mu\nu}\phi^{;\mu\nu} \cr 
&\qquad \qquad -\frac{1}{6}G_{5X}(\phi, X)\Biggl[(\Box \phi)^3 - 3\Box \phi_{;\mu\nu}\phi^{;\mu\nu} \cr
&\qquad \qquad \qquad \qquad \qquad \qquad + 2{\phi_{;\mu}}^{;\nu} {\phi_{;\nu}}^{;\lambda} {\phi_{;\lambda}}^{;\mu}  \Biggr] \,.\label{L5} 
\end{align}
Here $;_\mu$ is a covariant derivative and $X = -\phi_{;\mu}\phi^{;\mu}/2$, the canonical kinetic energy density of $\phi$.  The Lagrangian in Eq.~(\ref{fullLag}) is the most general Lagrangian density in the Horndeski theory. What we now focus on is a phenomenon occurring in Hubble time scale. In other words, we assume that gravity at small scale is irrelevant to our analysis at cosmological distance. This assumption is reasonable when testing GR at cosmological scale due to nontrivial screening mechanisms in a nonlinear regime known as Chameleon mechanisms \cite{Chameleon} or Vainstein mechanisms \cite{Vainstein}. Indeed, Babichev et al.  \cite{Babichev2011}  and Kimura et al. \cite{Vainsitein_cosmo2012} have successively reported that in the Horndeski theory the Vainstein mechanism recovers GR at a short distance, while time variation of gravitational coupling at cosmological distance remains unsuppressed. These facts indicate that we are justified to assume that a GW waveform is modified only at cosmological scale. As for a cosmological background, we define the flat Friedmann-Lema\^itre-Robertson-Waker (FLRW) metric as 
\begin{align}
ds^2 = -dt^2 + a^2(t)\delta_{ij}dx^idx^j\,.\label{FRW}
\end{align}
We now focus on the late-time of the universe below redshift $z = 1$. In this regime, time dependent functions are approximately given by Taylor expansion in powers of $H_0 t_{LB}$, where $H_0$  and $t_{LB}$ are the Hubble constant and the look back time given by
\begin{align}
& t_{LB}(z) = \int^{z}_0{\frac{dz'}{H(z') \cdot (1+z')}}\,, \label{t_LB} \\
&H(z) = H_0\sqrt{\Omega_{m0}(1+z)^3 + 1-\Omega_{m0}}\,.\label{Hz}
\end{align}
Here $H(z)$ is the Hubble parameter and $\Omega_{m0}$ is the matter density parameter. As we will see in Sec.~{\ref{ssec:setup}} we consider the two cases; $\Omega_{m0} = 0.308$ ($\Lambda$CDM) and $\Omega_{m0} = 1$ (Einstein-de-Sitter (EdS)) to give the Hubble parameter for each case. 
Now we expand the scalar field $\phi(t)$ as
\begin{align}
&\phi(t) \simeq M_{\rm pl} \left \{ a_0 + a_1H_0 t_{LB} + \frac{a_2}{2}(H_0 t_{LB})^2 \right \}\,, \label{phi} 
\end{align}
where $M_{\rm pl}$ is the Planck mass. The coefficients $a_n (n = 0,1,2)$ are arbitrary parameters with range $-1 < a_n < 1$. Practically in $z < 1$ it is enough to generate all models with these three parameters. 

Next we parameterize arbitrary functions $G_i (i = 2,3,4,5)$ in the Horndeski theory as  
\begin{align}
&G^{(\rm app)}_i (\phi, X) \equiv {\cal G}_i \Biggl \{\sum_{\rho = \hat{\phi}, \hat{X}}g_{i \rho}\rho + \sum_{\rho, \sigma = \hat{\phi}, \hat{X}}\frac{g_{i \rho \sigma}}{2}\rho \sigma  \cr
& \qquad \qquad \qquad \qquad \qquad + \sum_{\rho, \sigma, \lambda = \hat{\phi}, \hat{X}}\frac{g_{i \rho \sigma \lambda}}{6}\rho \sigma \lambda \Biggr \} \ \cr
& \qquad \qquad \qquad \qquad \qquad \qquad \qquad (i = 2,3,4,5)\,, \label{G_i}
\end{align}
where $\hat{\phi}$ and $\hat{X}$ are dimensionless quantities given as $\hat{\phi} \equiv \phi/M_{\rm pl}$ and $\hat{X} \equiv \dot{\phi}^2/2H^2_0M^2_{\rm pl}$. Throughout this paper, the dot is the derivative  with respect to $t$, not $t_{LB}$. Note that $dt_{LB} = -dt$. $g_{i \rho}$, $g_{i \rho \sigma}$, and $g_{i \rho \sigma \lambda}$ in Eq.~(\ref{G_i}) are the model parameters set at random ranging from -1 to 1.  ${\cal G}_i$ are normalization factors such as
\begin{align}
{\cal G}_2 = M^4\,,\ {\cal G}_3 =  \frac{M^3}{H_0^2}\,,\ {\cal G}_4 =  \frac{M^4}{H_0^2}\,,\ {\cal G}_5 = \frac{M^3}{H_0^4}\,,\label{calg_i}
\end{align}
 where $M \equiv \sqrt{M_{\rm pl}H_0}$. These normalization factors are determined in the way that the Lagrangian density of the system is the order of $M^2_{\rm pl}H^2_0 = M^4$. 
 This parameterization has in total 40 coefficients to distinguish models, which one is hardly able to obtain in analytical ways. This is an advantage of the numerical approach with the Monte Carlo method. Given the $\phi(t)$ and $G_i(t)$, we can compute all variables including $\alpha_M$ and $\alpha_T$ which are relevant to GW observations. In the Horndeski theory, $\alpha_M$ and $\alpha_T$ are given by 
\begin{align}
&M^2_*(t) \equiv 2 (G_4 -2XG_{4X} + XG_{5\phi} - \dot{\phi}HXG_{5X})\,, \label{M*Horn}\\
 &\alpha_M (t) = \frac{1}{HM^2_*}\frac{d M^2_*}{dt}\,, \label{aM_Horn}\\
&\alpha_T (t)= \frac{2X(2G_{4X} - 2G_{5\phi} - (\ddot{\phi} - \dot{\phi} H)G_{5X})}{M^2_*}\,, \label{aT_Horn}
\end{align} 
Substituting $\phi(t)$, the Hubble parameter in Eq.~(\ref{t_LB}), and all $G^{(\rm app)}_i$ in Eq.~(\ref{G_i}) into Eqs.~(\ref{M*Horn}) - (\ref{aT_Horn}), we can compute $\alpha_M$ and $\alpha_T$ as a function of redshift. 

However, one has to be careful about consistency of this formulation. According to our set-up above, we give the Hubble parameter as in Eq.~(\ref{Hz}) while we assume the time dependence of $\phi(t)$ without solving the Friedmann equations of the system, which may cause inconsistency of the Hubble parameter. To avoid this, we have to impose an alternative criterion to obtain proper solutions. In the following section, we concretely implement our method with a Monte Carlo analysis by imposing additional criteria and show that this remedy appropriately finds the consistent models.


\section{\label{sec:Horn_MC}Model classification with Monte Carlo simulation}
Next we classify models in the Horndeski theory into subgroups, depending on which the arbitrary functions $G_i$ play a role of accelerating the cosmic expansion. To this end, we now compute all physical quantities by randomly drawing all the coefficients from a uniform distribution, [-1, 1], with a Monte Carlo method. 

\subsection{\label{ssec:setup}Consistency and stability conditions}
\begin{figure}[ht]
 \begin{center}
   \includegraphics[width= 4.0cm]{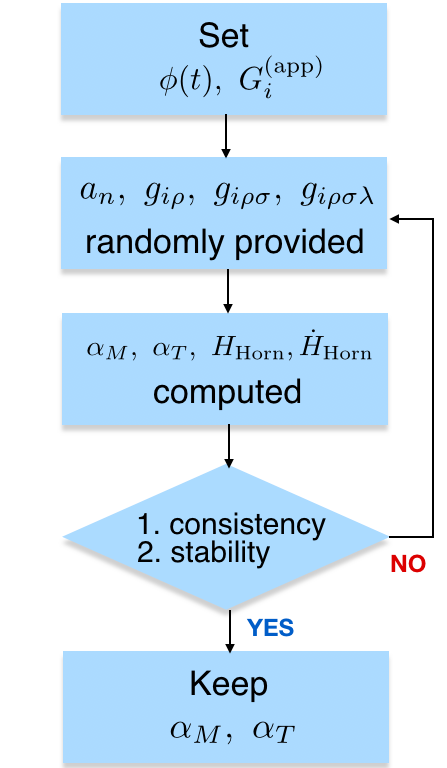}
  \caption{The procedure to extract observationally reliable models with Monte Carlo simulation}
    \label{flow1}
     \end{center}
\end{figure}

As shown in Fig.~\ref{flow1}, we filter the solutions by the following two conditions. 

As shown in Fig.~\ref{flow1}, here are two conditions introduced; consistency and stability

\begin{itemize}
\item 1. consistency : \\
Collecting the models whose cosmological time evolution, $H_{\rm Horn}$ and $\dot{H}_{\rm Horn}$.  $H_{\rm Horn}$ and $\dot{H}_{\rm Horn}$ are given by the Friedmann equations in Eqs.~(\ref{F1}) and (\ref{F2}) in Appendix.~\ref{app:computations}. To obtain $H_{\rm Horn}$ and $\dot{H}_{\rm Horn}$, we substitute $H_{\Lambda \rm CDM}$ and $\phi(t)$ for the right hand side of Eqs.~(\ref{E}) and (\ref{P}). Then we impose a consistency criteria as
\begin{align}
\biggl | 1- H_{\rm Horn}/H_{\rm \Lambda CDM} \biggr | < 20\%\,, \label{H20}\\
\biggl | 1- \dot{H}_{\rm Horn}/\dot{H}_{\rm \Lambda CDM} \biggr | < 20\%\,. \label{dH20} 
\end{align}
Eqs.~(\ref{H20}) and (\ref{dH20}) work to select the models that pass the observational bound on the cosmic expansion by assessing the deviation from $H_{\Lambda \rm CDM}$ and $\dot{H}_{\Lambda \rm CDM}$. We choose the allowed range of estimation errors for the Hubble parameter to 20$\%$ based on current variable observations of the Hubble parameter below $z = 0.1$ shown in the Table.~I of \cite{Hubble_Ratra}. 
\item 2. stability : \\Avoidance of ghost and gradient instabilities for the perturbations of scalar and tensor modes, 
\begin{align}
Q_s >0,\  c^2_s > 0, \ Q_ T> 0, \ c^2_T > 0\,.\label{stability}
\end{align} 
All the quantities are given by Eqs.~(\ref{Qs}) -(\ref{cT2}) in Appendix {\ref{app:computations}} respectively. For the computation, we substitute $H = H_{\Lambda \rm CDM}$, $\dot{H} = \dot{H}_{\Lambda \rm CDM}$ in the quantities. Matter density $\tilde{\rho}_m$ and pressure $\tilde{p}_m$ identified with cold dark matter density such as $\tilde{\rho}_m = \Omega_{m0}a^{-3}/M^2_*$ and $\tilde{p}_m = 0$ respectively.  The stability conditions guarantee the linear perturbation at cosmological scale. 
\end{itemize}

In addition to the conditions above, we assume the followings to make our discussion transparent. The third order coefficients of $G_2$ or $G_3$ are set to be zero because these are not directly related to all the physical quantities considered above. Then for simplicity the current value of the scalar field $\phi_0$ is set to be zero i.e. $a_0 = 0$, To satisfy the condition $\phi_0 = 0$, we have to restrict the function $G_2$ in a form of $G_2 = g_2 + g_{2X}X +g_{2\phi\phi}\phi^2 $. 

\subsection{\label{ssec:Models}Model distribution on the observables of gravitational waves}
 We show the distribution of models on $\alpha_T$ - $\alpha_M$ plane both of which are constrained from a GW measurement. In Sec.~\ref{ssec:setup}, we explained the procedure to assign the values of all the parameters. Executing this procedure provides all the physical quantities, including $\alpha_T$ and $\alpha_M$, at a referred redshift. We repeat the procedure and produce 1,000,000 discriminative models. In the following subsection, we firstly deal with an typical example $G_4,G_5 \neq 0, G_2 = 0 = G_3$, and then  we provide a general case that all functions including $G_2$ and $G_3$ are switched on.
 
\subsubsection{\label{sssec:G4G5}The effect of $G_4$ and $G_5$}
The functions $G_4$ and $G_5$ play significant roles for $\alpha_T$ and $\alpha_M$. In fact, one can see that $\alpha_T$ and $\alpha_M$ are determined solely by $G_4$ and $G_5$ in Eqs.~(\ref{aM_Horn}) and (\ref{aT_Horn}). In addition, $G_4$ and $G_5$ can control the cosmic accelerating expansion. Due to these importance, we firstly perform our simulation leaving $G_4$ and $G_5$ nontrivial while setting $G_2 = 0 = G_3$. 

Firstly we see how the different histories of cosmic expansion affects the model distribution on the $\alpha_T$ - $\alpha_M$ plane. Here we refer to the cosmic expansion in the EdS universe. We now obtain the distribution in the case of the EdS just by replacing $H_{\Lambda \rm CDM}$ and $\dot{H}_{\Lambda \rm CDM}$ in Eqs.~(\ref{H20}) and (\ref{dH20}) with those of EdS model, $H_{\rm EdS}$ and $\dot{H}_{\rm EdS}$, respectively. Figure 2 shows how distinctively the models distribute on the $\alpha_T$-$\alpha_M$ plane under two different histories of the cosmic expansion.
Moreover, to realize the cosmic expansion close to the case in the $\Lambda$CDM model with $G_4$ and $G_5$, either  $\alpha_T$ or $\alpha_M$ must be ${\cal O}(1)$. This result is expected from the analytic estimation in Lombriser and Taylor \cite{Lombriser2016}, but the shape of our distribution is different in detail from theirs.
The dots are very sparse at at the left top or right bottom where either $\alpha_T$ or $\alpha_M$ is extremely small. The main reason for this is due to the random sampling of models from whole possible models. In other words, the models which have tiny values of $\alpha_M$ and $\alpha_T$ need fine-tuning to realize the cosmic expansion in the same way as the $\Lambda$CDM model. 


\begin{figure}[ht]
  \begin{center}
   \includegraphics[width= 7.0cm]{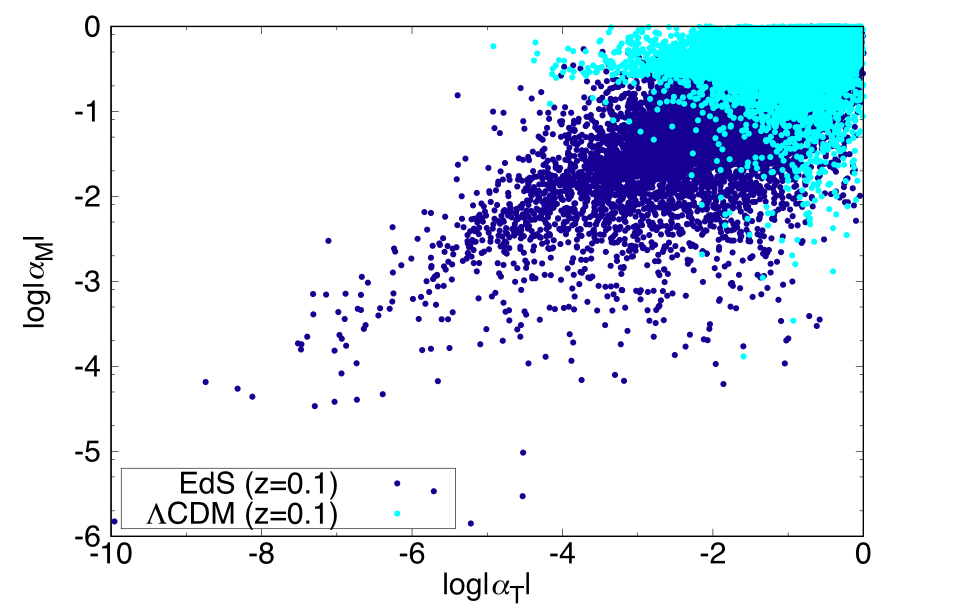}
  \caption{Distribution of the models in the $\alpha_T$-$\alpha_M$ plane with different cosmic expansion histories. The $\Lambda$CDM model (cyan dots) and the EdS model (dark-blue dots) are considered.}
    \label{fig1}
     \end{center}
\end{figure}
\subsubsection{\label{sssec:general}Model classification on the $\alpha_T$-$\alpha_M$ plane in general cases}
\begin{figure*}[htb]
  \begin{center}
     \includegraphics[width= 12.0cm]{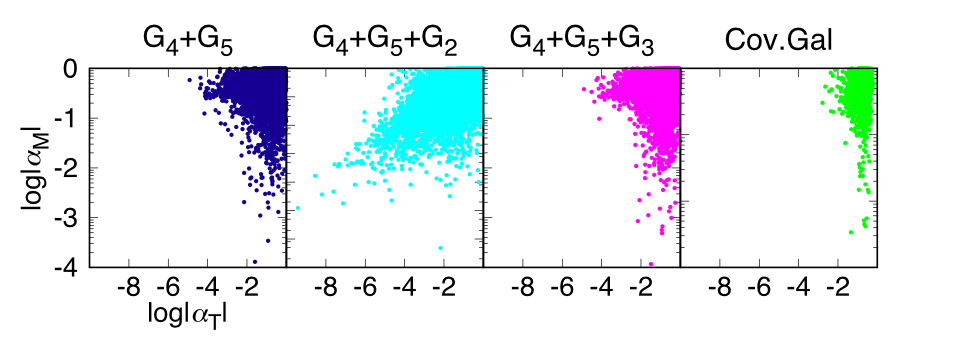}
  \caption{Distribution of models in each subclass shown in Table.~1 on $\alpha_T$ - $\alpha_M$ plane. }
    \label{fig2}
    \end{center}
\end{figure*}

We now allow all the model parameters to vary, namely $G_2$ and $G_3$ are both nonzero. The parameters in Eqs.~(\ref{phi}) and (\ref{G_i}) are now provided at random.  What we show in this section is how the models belonging to the different subclasses of the Horndeski theory are distributed on the $\alpha_T$-$\alpha_M$ plane. For instance, quintessence and the scalar field with nonlinear kinetic theory are exactly at the point of $\alpha_T = 0 = \alpha_M$, while $f(R)$ theory is on the $\alpha_M$ axis ($\alpha_T = 0$). We now classify the models into four categories shown in Table.~1. Based on the classification, we carry out the Monte Carlo simulation and obtain the distributions of each subclass in Fig.~\ref{fig2}.

\begin{table*}[htb]
\begin{tabular}{lccc}
\hline
Subclass of Horndeski theory & Parameters of $G^{(\rm app)}_i$ & Models & Refs. \\ \hline \hline
(I) $G_4 + G_5$ &$G_2, G_3 = 0$ & self acceleration & \cite{Lombriser2016}\\
 (II) $G_4+G_5 + G_2$ &$g_2, g_{2X}, g_{2\phi\phi} \neq 0$ & quintessence/nonlinear kinetic theory/ & \\
 & &$f(R)$ theories & \cite{Peebles-Ratra2003, Chiba2000, f(R)theories, LSSf(R)}\\
(III) $G_4+G_5+G_3$ & $G_3 \neq 0$ & cubic galileons & \cite{cubic_Gal, cubic_Gal_nonlinear}\\
(IV) Cov.Gal & $g_{2X}, g_{3X}, g_{4XX}, g_{5XX} \neq 0$ & covariant galileons &\cite{CovGal,Galileon} \\ \hline \hline
\end{tabular}
\caption{Division of subclasses with parameters in $G^{(\rm app)}_i$ and corresponding theories}
\end{table*}

As we see in Fig.~\ref{fig2}, the models except the subclass (II) are distributed in the domain with large $\alpha_T$. This is because in those cases $G_4$ and $G_5$ mainly drive the cosmic accelerating expansion and $\alpha_T$ consequently becomes large.  In the subclass (II), on the contrary, the models diagonally concentrate toward the area in which both $\alpha_T$ and $\alpha_M$ are small.  This is because $G_2$ in turn plays a role to accelerate the cosmic expansion, which relaxes the constraints on $G_4$ and $G_5$. In this subclass, we also find that the models are predicted to align along the diagonal line, 
$|\alpha_T| \propto |\alpha_M|^2$.  We discuss analytically why this feature appears in the following way.
First of all, we assume that $\dot{\phi}$ is initially tiny and the time evolution of $\dot{\phi}$ is very slow as $|\ddot{\phi}/H\dot{\phi}| \ll 1$. In this case,  we expand $G_i$ as
\begin{align}
&G_i(\phi, X) \simeq G_i(\phi_0, X_0) - G_{i\phi}(\phi_0, X_0) \dot{\phi}_{0}t_{LB} \,, \label{Gi_slow}
\end{align}
where the subscript $0$ denotes the values at the present time. Hereafter we simplify the expressions $G_i(\phi_0, X_0) = G_{i,0}$ and $G_{i \rho}(\phi_0, X_0) = G_{i \rho,0}$, where $\rho = \phi$ or $X$. 
With the same way as Eq.~(\ref{Gi_slow}), we obtain the observable parameters $\alpha_M$ and $\alpha_T$ from Eqs.~(\ref{M*Horn}) - (\ref{aT_Horn}) as
\begin{align}
&\alpha_M \simeq \frac{G_{4\phi,0}}{G_{4,0}}\dot{\phi}_0\,, \label{aM_app}\\
&\alpha_T \simeq \frac{2(G_{4\phi,0} - G_{5\phi,0})}{G_{4,0}}X_0\,,\label{aT_app}
\end{align} 
Considering $X_0 = \dot{\phi}_0^2/2$ we obtain the relation between $\alpha_M$ and $\alpha_T$ as
\begin{align}
\frac{\alpha_T}{\alpha^2_M} \simeq \frac{G_{4,0}(G_{4X,0} - G_{5\phi,0})}{{G_{4\phi,0}}^2} = \frac{g_4(g_{4X} - g_{5\phi})}{{g_{4\phi}}^2}\,.\label{am_at}
\end{align}
Eq.~(\ref{am_at}) is only valid unless $g_{4\phi} = 0$. The second equality is obtained from Eq.~(\ref{G_i}). Since in our computation the model parameters are given by constants at random,  
we see that models distribute along the line of $|\alpha_T| \propto |\alpha_M|^2$, which corresponds to a diagonal line on the $\alpha_T$-$\alpha_M$ plane in the logarithmic scale.


Our analysis also suggests the statement that the naive parameterization of $\alpha$s is generally incorrect. In literature, it is often assumed that the time evolution of all $\alpha$s are proportional to the energy density of dark energy, $\alpha = \Omega_{DE} \alpha_i$, where $\Omega_{DE} (t) \equiv \tilde{\cal E}/3H^2_0$ and $\alpha_i$ an initial value of $\alpha$ \cite{Maximal2014}. However, as Linder has pointed out recently, this assumption is not always correct in the case of $f(R)$ gravity \cite{Linder2017}. Our results also support this statement.  Here we only see the correlation between $\alpha_T$ and $\alpha_M$, but our technique is easily applicable to investigate to know the correlations among other parameters including $\alpha_K$ and $\alpha_B$. We will address this issue in the future publication.

\section{\label{sec:real_obs}Observational constraints on models}
As seen in the previous section, $G_4$ and $G_5$ induce the large deviation of $\alpha_T$ even in general cases. Next we derive the observational constraints on the Horndeski theory with GW propagation. Firstly we derive the analytic approximation for the observables of GW, which is applicable to the future observations of GW. Then we apply the expressions to the latest detection of GW170817 and GRB170817A to obtain the observational bound on the Horndeski theory.
  
\begin{figure}[ht]
  \begin{center}
  \includegraphics[width= 8.0cm]{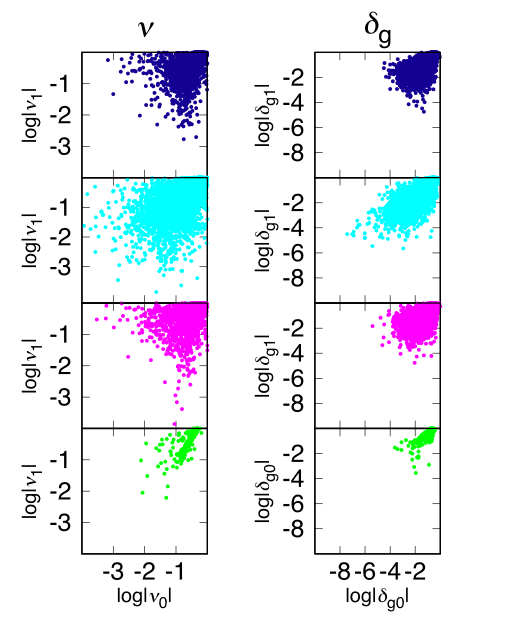}
  \caption{Model distributions on the observable parameter plane for the subclasses shown in Table.~1. Each color of the dots corresponds to that in Fig.~\ref{fig2}.} 
    \label{fig3}
     \end{center}
\end{figure}

\begin{figure}[ht]
 \begin{center}
  \includegraphics[width= 7.0cm]{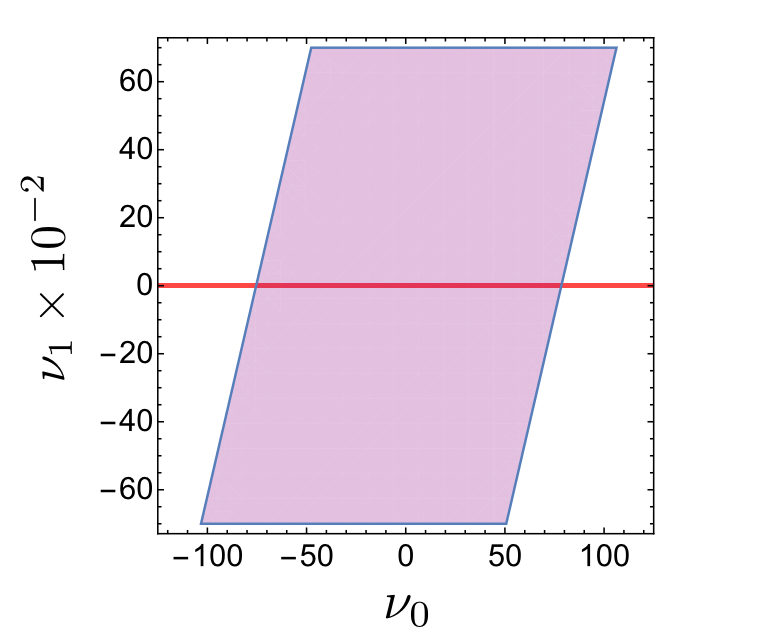}
  \caption{Observational constraint on $\nu_0$ and $\nu_1$. The width of the colored region is given in $1\sigma$ confidence level of the GW observation. The red solid line is $\nu_1 = 0$.} 
    \label{fig4}
   \end{center}
\end{figure}

From Eqs.~(\ref{calD})-(\ref{cT}), we can write down ${\cal D}$ and $\Delta T$ in terms of $\alpha_M$ and $\alpha_T$. Note that ${\cal D}$ and $\Delta T$ are the observables given after integrating all effects between emission and detection. However, we are now interested in the case that all the quantities vary in the cosmological time scale. In such a case it is justified to use the Taylor expansion with respect to $H_0t_{LB}$ as given in Eq.~(\ref{phi}). With the definitions of the time evolution of $\nu$ and $\delta_g$ as
\begin{align}
&\nu = \nu_0 - \nu_1H_0t_{LB}\,, \label{nu_def} \\
&\delta_g = \delta_{g0} - \delta_{g1}H_0t_{LB}\,, \label{dg_def}
\end{align}
expanding up to the next-to leading order in $H_0t_{LB}$ gives the approximated expressions of Eqs.~(\ref{calD}) and ~(\ref{Delta_T}) as
\begin{align}
&{\cal D} \simeq \frac{1}{2}\left \{\nu_0 \ln{(1+z)}- \frac{\nu_1}{2}(H_0t_{LB})^2 \right \} \,,\label{calD_obs}\\
&\Delta T \simeq \frac{1}{H_0} \left \{ \delta_{g 0}H_0t_{LB}- \frac{\delta_{g 1}}{2}(H_0t_{LB})^2 \right \}\,,\label{Delta_T_obs}
\end{align}
where the coefficients are related to $\alpha$ functions as
\begin{align}
&\nu_0 = \alpha_{M,0}\,,\label{nu0}\\
&\nu_1 = \frac{\dot{\alpha}_{M,0}}{H_0}\,,\label{nu1} \\
&\delta_{g 0} = -\frac{\alpha_{T,0}}{2}\,, \label{dg0}\\
&\delta_{g 1} = -\frac{\dot{\alpha}_{T,0}}{2H_0}\,.\label{dg1}
\end{align}
Converting the model parameters in the previous section to those in Eqs.~(\ref{nu0}) - (\ref{dg1}), we obtain the distributions of the observable parameters as shown in Fig.~\ref{fig3}. 
 Likely to Fig.~\ref{fig3}, Fig.~\ref{fig4} shows that the effect of the $G_4$ and $G_5$ functions is distinguishable in the both parameter $\nu$ and $\delta_g$, while Fig.~\ref{fig3} additionally shows the information about the time evolution of the models in terms of $\nu_1$ and $\delta_{g1}$. 
 
We now give observational constrains on $\nu_0$, $\nu_1$, $\delta_{g0}$, and $\delta_{g1}$ from the detection of GW170817 and GRB170817A. The reason to choose this GW event is that the redshift of the GW is independently measured from the optical follow-up observation of NGC4993 \cite{GW170817_multi}, which resolves the degeneracy between the redshift and the luminosity distance in the GW observation. 
For the observables, ${\cal D}$ and $\Delta T$, we have to take into account their estimation errors. In the case of measuring the arrival time difference, 
errors arise due to the accuracy of time resolution and intrinsic time delay at the source. As mentioned in the paper \cite{AN_TN_cT}, the time resolution is sufficient so that we ignore the timing error and consider only the arrival time difference. As we see in \cite{GW170817}, the arrival time difference is measured as 1.7s, which gives the upper bound on $c_T$. We also set the intrinsic time delay at the source to 10 sec to obtain the lower bound of $c_T$. 

\begin{figure}[ht]
\vspace{-0.5cm}
  \begin{center}
  \includegraphics[width= 7.0cm]{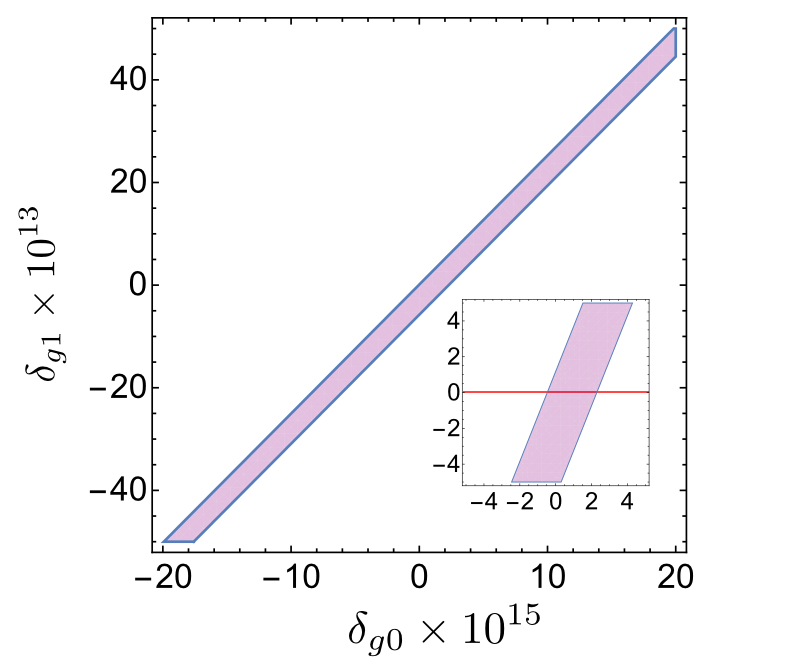}
  \caption{Observational constraint on $\delta_0$ and $\delta_1$. The width of the colored region is given between the lower and upper bounds. The small panel in the right bottom is the enlarged version around the center of the figure. he width of the colored region is given in $1\sigma$ confidence level of the GW observation. The red solid line is $\delta_{g1} = 0$. } 
    \label{fig5}
     \end{center}
\end{figure}

While the constraint on $\nu$ is given by comparison between the observed luminosity distance and the computed one using redshift determined by optical observations.
As shown in \cite{GW170817} the observed luminosity distance is given with error by $40^{+8}_{-14}$Mpc.  The computed luminosity distance at given redshift is obtained once we assume the cosmology.
To obtain the computed luminosity distance, $H_0$, and $t_{LB}$, we assume the cosmology to the best fit $\Lambda$CDM model in \cite{Planck2015} i.e. $H_0 = 67.8\rm km\ s^{-1} Mpc^{-1}$ and $\Omega_{m0} = 0.308$.  After all the procedures, we finally obtain the constraints as shown in Fig.~\ref{fig4} and Fig.~\ref{fig5}.
Comparing the left panel in Fig.~\ref{fig3} with Fig.~\ref{fig4}, we find that $\nu_0$ and $\nu_1$ are loosely constrained not enough to distinguish the models in Table.~I. By contrast, in the right panel in Fig.~\ref{fig3} and Fig.~\ref{fig5} all the models are excluded by the single observation of GW170817 unless one makes exceptional fine-tuning for the models. Consequently, we conclude that the models in the Horndeski theory which include $G_4$ or $G_5$ without $G_2$ can be excluded unless the models parameters are fine-tuned so that $c_T=1$. The models in the subclass (II) such that quintessence, nonlinear kinetic theory, or $f(R)$ theories survive because these models satisfy $c_T=1$. In other words, without fine-tuning for the $G_4$ and $G_5$ functions, the cosmic expansion must be driven by the $G_2$ function.

If we take the cases that $\nu_1 = 0$ or $\delta_{g1} = 0$, we obtain the constraints on $\nu_0$ and $\delta_{g0}$ as
\begin{align}
 -75.3 \leq &\nu_0 \leq 78.4\,,\label{nu0_obs}\\
-4.7\times10^{-16} \leq &\delta_{g0} \leq 2.2\times10^{-15}\,.\label{dg0_obs}
\end{align} 
$\nu_0$ is too loose to constrain the models in the Table.~I, while $\delta_{g0}$ is well determined enough to exclude the models. The constraint of $\delta_{g0}$ is consistent with the one in \cite{GW170817_cT}. 

Finally, we comment on a future prospect of tightening the constraints. In the above, we used a single source to constrain the Horndeski theory. However, given multiple sources at different distances in the future observation, one can tighten the constraint by combining the sources in two ways \cite{AN_GWpropagation}. First with a single source, there exists a parameter degeneracy in the time evolution, as shown by shaded bands in Figs. \ref{fig4} and \ref{fig5}. If one uses multiple sources at different distances, the bands cross at a point and the degeneracy is broken.
Second, currently the intrinsic time delay limits the sensitivity to $\delta_g$. However, the intrinsic time delay can be distinguished from modification of GW propagation 
and be partially canceled out by combining multiple signals.
This is possible because those effects depend on redshift differently at cosmological distances.

\section{\label{sec:Conclusions}Conclusions}
In this paper, we have discussed about constraints on Horndeski theory with GW propagation. We firstly reviewed the general framework for the waveform deformation from modified gravity including the Horndeski theory. Then we numerically formulated the Horndeski theory at cosmological scale to compute $\alpha_M$ and $\alpha_T$. 

Next, we performed the Monte Carlo simulation that keeps the models consistent with the observations of the cosmic expansion. In this procedure, we adopted two criteria; consistency and stability. Carrying out the simulation, we obtained the model distribution on $\alpha_T$-$\alpha_M$ plane.   Then we found that $\alpha_M$ and $\alpha_T$ have large values in the models including only $G_4$ or $G_5$, while including $G_2$ together with $G_4$ or $G_5$ allows both $\alpha_M$ and $\alpha_T$ to be smaller. Thanks to this feature, the models are 
significantly distinguishable depending on whether the models include the function $G_2$ or not.

Finally, we constrained the Horndeski theory from the simultaneous detection of GW170817 and GRB170817A. We 
provided the observational bounds for the physical parameters that involve in GW propagation ;$\nu_0$, $\nu_1$, $\delta_{g 0}$, and $\delta_{g 1}$. 
As a result, we found that the constraints on $\nu_0$ and $\nu_1$ are still too weak to distinguish the specific models shown in the Table.~I, while those on $\delta_{g 0}$, and $\delta_{g 1}$ are sufficiently strong enough to exclude the models that contain $G_4$ or $G_5$ without $G_2$. 
Consequently, we concluded that the model space of the Horndeski theory must be significantly reduced to explain the cosmic accelerating expansion and the GW propagation simultaneously. In other words, in the Horndeski framework the main driver of the cosmic accelerating expansion should be $G_2$. At present, the models such that quintessence, nonlinear kinetic theory, or $f(R)$ theories are still allowed.

In addition to our work, we comment on the theories other than the Horndeski theory including higher derivatives of a scalar curvature and a scalar field. Our formulation in effect contains those theories because the higher derivative terms become too tiny to be observed at cosmological scale. 
This argument agrees with the recent reports just after the detection of GW170817 was announced \cite{GW170817_DE_Creminelli, GW170817_Jeremy, GW170817_DE_Jose, GW170817_Baker}. However, our work quantitatively discusses how much the fine-tuning of the model is required within the current observational errors. In addition, as shown in Eq.~(\ref{nu0_obs}), we obtained for the first time the constraint on the amplitude damping parameter $\nu$ by the observation of GW170817. Although the constraint is loose, it plays an important role to restrict further the models whose $\delta_g$ is fine-tuned to zero. 

In the end, we report on an accidental finding
 that the parameterization of $\alpha \propto \Omega_{DE}\alpha_i$ is not valid in general. This assumption is now widely used when computing the cosmological observables particularly CMB angular power spectrum by using Einstein-Boltzmann solvers \cite{EFTCAMB, hiCLASS, EBcodes2017}. 
Therefore, it is important to revisit the previous constraints on the Horndeski theory parameterized $\alpha \propto \Omega_{DE}\alpha_i$ and to investigate the application of our simulation to other cosmological observations such as CMB or LSS. We will address this issue in future publication.

\begin{acknowledgments}
S.A is supported by a Grant-in-Aid for Japan Society for the Promotion of Science Research under
Grants No.17J04978. This work was supported by JSPS KAKENHI Grant Number JP17H06358.
\end{acknowledgments}

\appendix
\section{\label{app:computations}Computation of model parameters}
We introduce physical quantities necessary for computation in the main text.  Here we use the $\alpha$ parametrization in Bellini and Sawacki \cite{Maximal2014}. First of all, the time-evolving the parameters are defined as
\begin{align}
&M^2_* \equiv 2 (G_4 -2XG_{4X} + XG_{5\phi} - \dot{\phi}HXG_{5X})\,,\\
&HM^2_*\alpha_M \equiv \frac{d}{dt}M^2_*\,, \label{aM_app}\\
&H^2M^2_*\alpha_K \equiv 2X(G_{2X}+2XG_{2XX}-2G_{3\phi}-2XG_{3\phi X}) \cr
&\qquad \qquad +12\dot{\phi}XH(G_{3X}+XG_{3XX}-3G_{4\phi X}-2XG_{4\phi XX}) \cr
&\qquad \qquad +12XH^2(G_{4X}+8XG_{4XX}+4X^2G_{4XXX}) \cr
&\qquad \qquad -12XH^2(G_{5\phi}+5XG_{5\phi X}+2X^2G_{5 XXX}) \cr
&\qquad \qquad + 4\dot{\phi}XH^3(3G_{5X}+7XG_{5XX}+2X^2G_{5XXX})\,, \label{aK_app} \\
&HM^2_*\alpha_B \equiv 2\dot{\phi}(XG_{3X}-G_{4\phi}-2XG_{4\phi X})\cr
&\qquad \qquad + 8XH(G_{4X}+2XG_{4XX}-G_{5\phi} -XG_{5\phi X}) \cr
&\qquad \qquad + 2\dot{\phi}XH^2(3G_{5X}+2XG_{5XX})\,, \label{aB_app}\\
&M^2_*\alpha_T \equiv 2X(2G_{4X} - 2G_{5\phi} - (\ddot{\phi} - \dot{\phi} H)G_{5X})\,\label{aT_app}.
\end{align}
Of all the four parameters, $\alpha_M$ and $\alpha_T$ are relevant to GW propagation. While, $\alpha_K$ and $\alpha_B$
are irrelevant to GW propagation, but they are necessary to evaluate the stability condition discussed in Sec.~\ref{ssec:setup}. The the Friedman equations in the Horndeski theory are given by
\begin{align}
&3H^2 = \tilde{\rho}_m + \tilde{\cal E}\,, \label{F1}\\
&2\dot{H} + 3H^2 = -\tilde{p}_m - \tilde{\cal P}\,, \label{F2}
\end{align}
where $\tilde{\rho}_m \equiv \rho_m/M^2_*$ and $\tilde{p}_{m} \equiv p_m/M^2_*$. Then $\tilde{\cal E}$ and $\tilde{\cal P}$ are given by
\begin{align}
&M^2_*\tilde{\cal E} \cr
&= -G_2+2X(G_{2X}-G_{3\phi}) \cr 
&+6\dot{\phi} H(XG_{3X}-G_{4\phi}-2XG_{4\phi X})\cr
&+12H^2X (G_{4X}+2XG_{4XX}-G_{5\phi}-XG_{5\phi X})\cr
&+4\dot{\phi}H^3X(G_{5X}+XG_{5XX})\,, \label{E}\\
&M^2_*\tilde{\cal P} \cr
&= G_2-2X(G_{3\phi}-2G_{4\phi\phi})\cr
&+4\dot{\phi}H(G_{4\phi}-2XG_{4\phi X} + XG_{5\phi\phi})\cr 
&-M^2_*\alpha_B H\frac{\ddot{\phi}}{\dot{\phi}}-4H^2X^2G_{5\phi X} + 2\dot{\phi}H^3XG_{5X}\,.\label{P}
\end{align}
As we can see in all the quantities above, the third derivatives of $G_2$ and $G_3$  implicitly affect on the quantities. Therefore, we set the third derivatives of $G_2$ and $G_3$ to be zero in the main text.\\
Finally, we provide essential quantities to avoid the ghost and gradient instability. The action at quadratic order of a scalar field $\zeta$ and tensor modes $h_{ij}$ are given by  
\begin{align}
&S_2 = \int{dtd^3xa^3}\Biggl [Q_s \left(\dot{\zeta}^2-\frac{c^2_s}{a^2}(\partial_i \zeta)^2\right) \cr 
& \qquad \qquad \quad \quad+ Q_T \left(\dot{h}_{ij}^2-\frac{c^2_T}{a^2}(\partial_k h_{ij})^2\right)  \Biggr]\,, \label{S2}
\end{align}
where 
\begin{align}
&Q_s = \frac{2M^2_*D}{(2-\alpha_B)^2}, \label{Qs}\\
&c^2_s = -\frac{(2-\alpha_B)}{H^2D}\Biggl[\dot{H}-\frac{1}{2}H^2\alpha_B(1+\alpha_T)\cr
&\quad \quad  -H^2(\alpha_M-\alpha_T)-H\dot{\alpha}_B+\tilde{\rho}_m + \tilde{p}_m \Biggr]\,,\label{cs2}\\
&D \equiv \alpha_K + \frac{3}{2}\alpha^2_B\,,\nonumber
\end{align}
while 
\begin{align}
&Q_T = \frac{M^2_*}{8}\,,\label{QT}\\
&c^2_T = 1+\alpha_T\,,\label{cT2}
\end{align}
To avoid the theoretical instabilities, we should impose the condition that $Q_s > 0$, $c^2_s > 0$, $Q_T > 0$ and $c^2_T >0$.

\end{document}